\documentclass[12pt]{article}

\usepackage{times}
\usepackage{epsfig}
\usepackage{amssymb}
\usepackage{subfigure}
\usepackage{graphicx}
\usepackage{color}
\usepackage{jheppub} 

\begin{document}

\newcommand{\nc}{\newcommand}
\newcommand{\rnc}{\renewcommand}

\rnc{\baselinestretch}{1.24}    
\setlength{\jot}{6pt}       
\rnc{\arraystretch}{1.24}   

\makeatletter
\rnc{\theequation}{\thesection.\arabic{equation}}
\@addtoreset{equation}{section}
\makeatother


 \renewcommand{\thefootnote}{\fnsymbol{footnote}}

\newcommand{\tcb}{\textcolor{blue}}
\newcommand{\tcr}{\textcolor{red}}
\newcommand{\tcg}{\textcolor{green}}


\def\be{\begin{eqnarray}}
\def\ee{\end{eqnarray}}
\def\nn{\nonumber\\}


\def\ct{\cite}
\def\la{\label}
\def\eq#1{\eqref{#1}}


\def\a{\alpha}
\def\b{\beta}
\def\g{\gamma}
\def\G{\Gamma}
\def\d{\delta}
\def\D{\Delta}
\def\e{\epsilon}
\def\et{\eta}
\def\ph{\phi}
\def\Ph{\Phi}
\def\ps{\psi}
\def\Ps{\Psi}
\def\k{\kappa}
\def\l{\lambda}
\def\L{\Lambda}
\def\m{\mu}
\def\n{\nu}
\def\th{\theta}
\def\Th{\Theta}
\def\r{\rho}
\def\s{\sigma}
\def\S{\Sigma}
\def\ta{\tau}
\def\o{\omega}
\def\O{\Omega}
\def\pr{\prime}


\def\half{\frac{1}{2}}
\def\goto{\rightarrow}

\def\na{\nabla}
\def\grad{\nabla}
\def\curl{\nabla\times}
\def\div{\nabla\cdot}
\def\pa{\partial}
\def\fr{\frac}

\def\bra{\left\langle}
\def\ket{\right\rangle}
\def\lb{\left[}
\def\lc{\left\{}
\def\ls{\left(}
\def\lp{\left.}
\def\rp{\right.}
\def\rb{\right]}
\def\rc{\right\}}
\def\rs{\right)}

\def\vac#1{\mid #1 \rangle}


\def\td#1{\tilde{#1}}
\def\check{ \maltese {\bf Check!}}


\def\Tr{{\rm Tr}\,}
\def\det{{\rm det}}
\def\text#1{{\rm #1}}


\def\bc#1{\nnindent {\bf $\bullet$ #1} \\ }
\def\ch {$<Check!>$ }
\def\ss {\vspace{1.5cm}}
\def\inf{\infty}

\begin{titlepage}

\hfill\parbox{2cm} { }

 
\vspace{2cm}

\begin{center}
{\Large \bf Holographic RG flow triggered by gluon condensate}

\vskip 1. cm
   {Chanyong Park$^{a}$\footnote{e-mail : cyong21@gist.ac.kr}}

\vskip 0.5cm

{\it $^a$ Department of Physics and Photon Science, Gwangju Institute of Science and Technology\\ 
 Gwangju  61005, Korea}

\end{center}

\thispagestyle{empty}

\vskip3cm


\centerline{\bf ABSTRACT} 


\vspace{0.5cm}

By applying the holographic method, we study a non-perturbative renormalization group (RG) flow triggered by a gluon condensate. After introducing a bulk scalar field in an AdS space related to the gluon condensate, we investigate the trace anomaly proportional to the gluon condensate. The holographic calculation reproduces the one-loop trace anomaly known in the lattice QCD. We also show that higher loop corrections give rise to additional contributions and modify the one-loop trace anomaly. \\

"Contribution to the Proceedings of the 6th International Conference on Holography, String Theory and Spacetime, Da Nang, Vietnam, February 2023".

\vspace{2cm}

\end{titlepage}

\renewcommand{\thefootnote}{\arabic{footnote}}
\setcounter{footnote}{0}



\section{Introduction}

After the proposal of the AdS/CFT correspondence (or holography) \cite{Maldacena:1997re,Gubser:1998bc,Witten:1998qj,Witten:1998zw}, there were many attempts to understand strongly interacting systems by using the holographic method. One of the important features in the holography is that non-perturbative quantum nature of a quantum field theory (QFT) can be realized by a one-dimensional higher classical gravity. In this case, a dual QFT lives at the boundary of a bulk geometry and the boundary position is reinterpreted as the energy scale observing the dual QFT. In the holographic studies, therefore, we can investigate the energy scale dependence of QFT nonpertubvatively by varying the boundary position \cite{Henningson1998,Henningson2000,Freedman1999,Gubser:1999vj,deBoer:1999tgo,Verlinde:1999xm,Boer2001,Skenderis:1999mm,Polchinski:2011im,Papadimitriou:2010as,Papadimitriou:2004ap,Skenderis2002,Heemskerk:2010hk,Erdmenger:2020hug} or by increasing the system size \cite{Kim:2016jwu,Park:2018ebm,Park:2019pzo,Park:2020nvo,Park:2022mxj}. The energy scale dependence of a QFT is called a renormalization group (RG) flow. In QFT, in general, it is very hard to find a non-perturbative RG flow exactly because of infinitely many higher loop corrections. However, the AdS/CFT correspondence may provide us a new chance to obtain information about the non-perturbative RG flow of the dual QFT. In this work, we will holographically investigate the trace anomaly triggered by the gluon condensate in QCD \cite{DiGiacomo:1981lcx,Trinchero:2013joa}. This proceeding paper is based on Ref. \cite{Park:2021nyc}. 

When a marginal operator deforms a conformal field theory (CFT), the deformed theory still remains conformal at the tree level. This feature can be explained by a tree-level $\b$-function \cite{Hollowood:2009eh}
\be
\b_{tree} = - (d -\D) \l ,
\ee
where $\l$ means a coupling constant. For a marginal operator whose conformal dimension is given by $\D=d$, the tree-level $\b$-function automatically vanishes, $\b_{tree}=0$. This indicates that a deformed theory is independent of the energy scale due to the scale invariance. When we consider quantum corrections, this is not the case because quantum corrections can provide nontrivial contribution to the $\b$-function. In the UV region, the modified $\b$-function is expressed as
\be
\b = \b_{tree} + \b_{q}  ,
\ee 
where $\b_q$ means small quantum corrections. At the quantum level, therefore, a tree-level marginal operator depending on the sign of $\b_q$ changes into one of a marginally relevant, truly marginal and marginally irrelevant operators. In this case, a marginally relevant operator gives rise to a nontrivial RG flow along which a UV CFT flows into a new IR theory.

It has been well known that QCD is asymptotic free. This means that QCD becomes a CFT at a UV fixed point. The trace of the stress tensor for a CFT automatically vanishes, $\bra {T^{\m}}_{\m} \ket =0$. However, if QCD has a nonvanishing gluon condensate, it can have a nonvanishing trace. This is  because the gluon condensate is a marginally relevant operator. In the lattice QCD, the trace of the stress tensor at the one-loop level is given by \cite{DiGiacomo:1981lcx,Trinchero:2013joa}
\be
\bra {T^{\m}}_{\m} \ket = - \fr{N_c}{8 \pi} \fr{\b_\l}{\l^2} \bra  G \ket ,
\ee
where $\l$ is a 't Hooft coupling constant. This was known as the trace anomaly triggered by the gluon condensate \cite{DiGiacomo:1981lcx,Trinchero:2013joa,Kim:2007qk,Kim:2008ax,Ko:2009jc}. In this work, we take into account the holographic dual of the gluon condensate and investigate its holographic RG flow holograhically.


\section{Holographic renormalization group flow}

The AdS/CFT correspondence has claimed that a strongly interacting $d$-dimensional CFT has a one-to-one map to a gravity theory in a $(d+1)$-dimensional AdS space. In this case, a CFT lives at the boundary of a bulk AdS space and the boundary position is reinterpreted as the energy scale observing the dual CFT. For example, a five-dimensional Euclidean AdS space is governed by the following action
\be			
S_{AdS} =  - \fr{1}{2 \k^2} \int d^{5} X \sqrt{g} \ls {\cal R}  - 2 \L  \rs + \fr{1}{\k^2} \int_{\pa {\cal M}} d^4 x \sqrt{\g} \ K .
\ee
Here the last term called the Gibbons-Hawking term is required to obtain a well-defined an equation of motion. A metric satisfying the equation of motion is given by an AdS geometry
\be
ds^2 = \fr{R^2}{z^2} \ls dz^2 + \d_{ij} \,  dx^i dx^j \rs ,
\ee
where $R$ is an AdS radius and $i, \, j$ run from $1$ to $4$. This gravity solution is the dual of a four-dimensional CFT whose $\b$-function automatically vanishes. 

In order to investigate a QFT depending on the energy scale nontrivially, we have to modify the bulk action by adding appropriate bulk fields. We take into account the following scalar field deformation
\be			\la{act:EuclEins-scalar}
S= S_{AdS} + \fr{1}{2 \k^2} \int d^{5} X \sqrt{g}   \ls \fr{1}{2} g^{MN}
\pa_M \ph \pa_N \ph   +  \half \fr{m^2}{R^2}  \ph^2 \rs    ,
\ee
which is the dual of a deformed CFT by a scalar operator. The gravitational backreaction of the scalar field  modifies the background AdS space, which is reinterpreted as a nontirival RG flow caused by a scalar operator on the dual field theory side. In the asymptotic region ($z \to 0$), the scalar field allows the following perturbative expansion 
\be
\ph (z) = c_1 z^{4-\D}  \ls 1 + \cdots \rs  + c_2 z^\D  \ls 1 + \cdots \rs ,   \la{Result:pscalar}
\ee
with
\be
\D = 2 +\sqrt{4 + m^2} ,   \la{Result:ConformalDim}
\ee 
where $c_1$ and $c_2$ are two integral constants. On the dual field theory side, $c_1$ and $c_2$  are reinterpreted as a source (or coupling constant) and vev of a deformation operator having the conformal dimension $\D$ \cite{Gubser:1998bc,Witten:1998qj,Witten:1998zw}. With the gravitational backreaction of $\ph$, the resulting metric has the following form
\be
ds^2 = \fr{R^2}{z^2} \ls dz^2 + f(z) \, \d_{ij} \,  dx^i dx^j \rs ,
\ee
where $f(z)$ has to satisfy equations for the scalar field and metric simulataneously.

From the CFT point of view, $\D=4$ with $m^2=0$ corresponds to the conformal dimension of a marginal operator, which does not modify the CFT at the tree level. On the contrary, an operator with $0< \D < 4$ is called a relevant operator which deforms a CFT and generates a new IR theory even at the tree level. This feature becomes manifest when we consider a $\b$-function. At the tree level, a $\b$-function is given by \cite{Hollowood:2009eh}
\be
\b_{tree}  \equiv \m \fr{\pa \l}{\pa \m}= - (d-\D) \l  ,    \la{Result:treebeta}
\ee
where $\l$ and $\m$ indicate a coupling constant and energy scale, respectively. If we further take into account quantum corrections, a $\b$-function in the UV region modifies into
\be
\b_\l =  - (d-\D) \l  + \b_q  ,   \la{Relation:bfunction}
\ee
where $\b_q$ means small quantum corrections. It is worth to note that this decomposition is possible only in the UV region. In the IR region, a general $\b$-function becomes non-perturbative, so that we cannot clearly distinguish tree and quantum contributions. Due to this reason, from now on we focus on the Einstein-scalar gravity in the UV region and compare it with a CFT deformed by the gluon condensate.

In order to realize the general $\b$-function \eq{Relation:bfunction} in the dual gravity theory, we introduce a normal coordinate $z= R e^{-y/R}$. Then, the AdS metric is rewritten as
\be		
ds^2 = dy^2 + e^{2 y/R} \d_{ij} \,  d x^{i} dx^{j}  .
\ee
The normal coordinate is convenient because the scale transformation of a field theory is represented as the translation in the $y$-direction. Assuming that the AdS boundary is located at $y=\bar{y}$, the energy scale of the dual CFT is expressed as $\m = e^{\bar{y}/R}/R$. When an AdS space is deformed by a scalar field, its gravitational backreaction modifies the metric into the following form
\be	
ds^2 = dy^2 + e^{2 A(y)}  \d_{ij} \,  d x^{i}    dx^{j}  ,    \la{Metric:deformed}
\ee
where $A(y)$ is a function of $y$. In this case, the energy scale of the dual field theory is defined as $\m = e^{A(\bar{y})}/R$. Now, we discuss the connection between the bulk field and a deformation operator of the dual field theory. The bulk field, as mentioned in \eq{Result:pscalar}, allows the perturbative expansion in the asymptotic region, which corresponds to a UV region on the dual field theory side. In this UV region, a coupling constant and vev of an operator are distinguishable. However, they are indistinguishable in the IR region (or in the interior region of the dual geometry) because of mixing of them. Therefore, it becomes unclear what a coupling constant is at the low energy scale. To overcome this issue, we identify the bulk scalar field $\ph(\bar{y})$ at the boundary with a coupling constant $\l(\m)$ relying on the energy scale $\m$. We will show later that this identification is consistent with identifying $c_1$ and $c_2$ with a coupling constant and vev in the UV region.

On the gravity side, in general, the scalar field and metric are governed by second-order differential equations
\be
0 &=& 24 \dot{A}^2 - \dot{\ph}^2 + 4 \L  + \fr{m^2}{R^2}  \ph^2 , 	\la{eq:consteq} \\
0 &=& 12 \ddot{A} + 24 \dot{A}^2 +   \dot{\ph}^2 + 4 \L + \fr{m^2}{R^2}  \ph^2  \la{eq:Adynamics} , \\
0 &=& \ddot{\ph} + 4 \dot{A} \dot{\ph}  ,  \la{eq:phdynamics}
\ee
where the dot means a derivative with respect to $y$. In this case, the first equation is a constraint and the others describe dynamics of the metric and scalar field. On the contrary, the RG equation is usually described by the first-order differential equation. To relate the bulk equations to the RG equation, we need to rewrite the bulk equations as the first-order forms. To do so, we apply the Hamilton-Jacobi formulation \cite{deBoer:1999tgo,Verlinde:1999xm,Skenderis:1999mm,Boer2001,Papadimitriou:2010as,Papadimitriou:2004ap}.  Let us consider the following general metric
\be			\la{met:admdecomp}
ds^2  
&=& N^2 dy^2  + \g_{ij}(y)  \, dx^{i}   dx^{j}  ,
\ee
where $N$ and $\g_{\m\n}$ are a lapse function and induced metric on a hypersurface at a given $y$. If $N=1$ and $\g_{ij}(y)= e^{2 A(y)} \d_{ij}$, this general metric reduces to the previous one in \eq{Metric:deformed}. Note that, since the lapse function plays a role of a Lagrange multiplier, we can set $N=1$, without loss of generality, after all cacluation. Using this general metric, the action \eq{act:EuclEins-scalar} can be rewritten as a functional of the extrinsic curvature \cite{deBoer:1999tgo,Verlinde:1999xm,Boer2001}
\be
S = \int d^{4} x d y \sqrt{\g} \ {\cal L}  ,
\ee
with
\be
{\cal L}  =  \fr{1}{2 \k^2} \lb N \ls - {\cal R}^{(4)} +K_{ij} K^{ij} - K^2 + 2 \L \rs
+ \fr{1}{2N}    \dot{\ph}^2   + \half \fr{m^2}{R^2}  \ph^2  \rb ,
\ee
where the extrinsic curvature is given by
\be
K_{ij} = \fr{1}{2 N} \fr{\pa \g_{ij}}{\pa y} .
\ee

Now, we assume that the bulk geometry has a flat boundary at $y=\bar{y}$. Then, the bulk geometry extends only to $-\infty < y \le \bar{y}$ and $\g_{ij} (\bar{y})$ becomes an induced metric of the boundary. In this case, since we assume the flat boundary, the curvature scalar on the boundary $ {\cal R}^{(4)}$ becomes zero. After introducing the canonical momenta of $\ph$ and $\g_{ij}$, the above action is rewritten as the first-order form
\be
S= \int d^{4} x d y \sqrt{\g} \ \ls \pi_{ij} \dot{\g}^{ij} + \pi_{\ph} \dot{\ph}  - N {\cal H}
\rs ,
\ee
where the canonical momenta are given by
\be			\la{def:conjugatemomenta}
\pi_{ij} 
&=& - \fr{1}{2 \k^2}  \ls K_{ij} - \g_{ij} K \rs , \nn
\pi_{\ph} 
&=& \fr{1}{2  \k^2} \dot{\ph}   .
\ee
Here, the Hamiltonian corresponds to a generator of the translation in the $y$-direction
\be
{\cal H} = 2 \k^2 \ls \g^{ik} \g^{jl} \pi_{ij} \pi_{kl} - \fr{1}{3} \pi^2 + \fr{1}{2} \pi_{\ph}^2 \rs  - \fr{\L}{\k^2}  -   \fr{1}{4} \fr{m^2}{R^2}  \ph^2   ,   \la{Result:HamCon}
\ee
where $\pi = \g^{ij} \pi_{ij}$, and the variation with respect to the lapse function leads to a Hamiltonian constraint, ${\cal H}=0$. When the Hamiltonian constraint is satisfied, varying the bulk action reduces to the variation of the boundary action 
\be
\d S_B =  \int_{\pa {\cal M}} d^4 x \sqrt{\g} \ls \pi_{ij}  \d {\g}^{ij} + \pi_{\ph} \d {\ph} \rs ,
\ee
where $\pa {\cal M}$ indicates the boundary at $y=\bar{y}$. According to the AdS/CFT correspondence, this boundary action $S_B$ can be regarded as a generating functional of the dual field theory. 

It is worth noting that the above boundary action even for the AdS space with $A(\bar{y}) = \bar{y}/R$ suffers from a divergence when $\bar{y} \to \infty$. Recalling that the limit of $\bar{y} \to \infty$ corresponds to a UV limit of the dual field theory, the divergence at $\bar{y}  \to \infty$ is associated with a UV divergence of the dual field theory which must be renormalized by adding an appropriate counterterm. Denoting a counterterm as 
\be
S_{ct} = - \fr{1}{2 \k^2} \int d^4 x  \sqrt{\g} \ {\cal L}_{ct}  ,
\ee 
the renormalized generating functional is given by
\be
\G [ \g_{\m\n}, \ph; \bar{y}]= S_B - S_{ct} ,    \la{Result:Gfunctional}
\ee
where $ \g_{\m\n}$ and $\ph$ are the values at the UV cutoff. In this setup, since the UV cutoff is artificial, the generating functional must be independent of $\bar{y}$ 
\be
0 = \fr{d \G}{d \bar{y}} .
\ee
This is also true at any scale of $\bar{y}$. Reinterpreting $\bar{y}$ as a RG scale, the generating functional  independent of the RG scale leads to the following RG equation 
\be
0 
= \m \fr{\pa \G}{\pa \m}  + \fr{\pa \g^{ij}}{\pa \log \m}  \fr{\pa \G}{\pa \g^{ij}} +  \fr{\pa \ph}{\pa \log \m}  \fr{\pa \G}{\pa \ph}  ,
\ee
where $\m=e^{A (\bar{y})}/R$ denotes the RG scale of the dual field theory. Here, we regards the boundary metric as a coupling constant coupled to the energy-momentum tensor. This prescription is useful to study the conformal anomaly of a QFT defined on a curved spacetime \cite{Friedan:1980aa,FRIEDAN1985318}. 

Using the following relation
\be
 \fr{d \g^{ij}}{d \log \m}  = - 2 \g^{ij} ,
\ee
the RG equation can be rewritten as the usual form
\be		\la{res:hRGeq}
0 = \fr{\m}{\sqrt{\g}} \fr{\pa \G}{\pa \m}  + \g^{ij} \bra T_{ij} \ket + \b_\ph  \bra O \ket  ,
\ee
where the $\b$-function and vev of the operator are defined as
\be
\b_\ph  &\equiv&  \fr{\pa \ph}{\pa \log \m} ,\\
\bra T_{ij} \ket &\equiv& - \fr{2}{\sqrt{\g}} \fr{\pa \G}{\pa \g^{ij} }  =  -  \ls 2 \pi_{ij} - \fr{1}{2 \k^2} \g_{ij} {\cal L}_{ct} \rs , \la{Result:vevofT} \\
\bra  O \ket &\equiv&    \fr{1}{\sqrt{\g}}  \fr{ \d \G}{\d \ph}  = \pi_\ph  + \fr{1}{2 \k^2} \fr{\pa {\cal L}_{ct} }{\pa \ph}  .  \la{Result:vevofoperator}
\ee
Before closing this section, there are several remarks. The RG equation in \eq{res:hRGeq} is the generalization of the QFT's RG equation. When the boundary metric is fixed, the generalized RG equation reproduces the QFT's result
\be		 
0 = \fr{\m}{\sqrt{\g}} \fr{\pa \G}{\pa \m}  + \b_\ph  \bra O \ket  .
\ee
Above, the generating functional in \eq{Result:Gfunctional} depends only on the coupling constants. In this case, the vev of the deformation operator is determined as a derivative of the generating functional with respect to the coupling constant, as shown in \eq{Result:vevofT} and \eq{Result:vevofoperator}.

\section{Holographic RG flow caused by gluon condensate}

For a $SU(N_c)$ Yang-Mills theory, the kinetic term of a gauge field is given by
\be
S_{YM} = - \fr{1}{4  g_{YM}^2}  \int d^4 x \sqrt{\g}   \  \  \Tr  F^2  .  \la{Action:YM}
\ee
If the vev of $G = - \Tr F^2$ does not vanish, it was known that the gluon condensate $\bra G \ket $ generates a trace anomaly at the one-loop level \cite{DiGiacomo:1981lcx,Trinchero:2013joa}
\be
\bra {T^{\m}}_{\m} \ket = - \fr{N_c}{8 \pi } \fr{\b_\l}{\l^2} \bra G \ket ,   \la{Result:GRGfloweqQCD}
\ee
where $\l = N_c \, g_{YM}^2$ is the 't Hooft coupling constant and $\b_\l$ means the $\b$-function of $\l$. At the tree level, since $G$ is a marginal operator with $\D=4$, the tree-level $\b$-function automatically vanishes. Therefore, the trace anomaly in \eq{Result:GRGfloweqQCD} corresponds to the quantum effect. Now, we describe the trace anomaly in the holographic dual gravity which may provide more information about a non-perturbative RG flow. In order to describe the gluon condensate holographically, we need to introduce a massless scalar field, $\ph$ with $m^2=0$, because it is the dual of a marginal operator with $\D=4$ according to \eq{Result:ConformalDim}. In order to identify the dual operator of $\ph$ with the gluon condensate ($O  =  G$), the value of $\ph$ at the boundary has to be identified with the coupling constant $\ph=1/(4  g_{YM}^2) = N_c / (4 \l ) $ in \eq{Action:YM}.

Introducing a superpotential satisfying for convenience \cite{Freedman:1999gp,Skenderis:1999mm,DeWolfe:1999cp,Csaki:2000wz, Gubser:1999pk, Kehagias:1999tr,Csaki:2006ji}
\be		\la{ans:superpotential}
W(\ph) = 6  \dot{A}  .
\ee
the bulk equations in \eq{eq:consteq}, \eq{eq:Adynamics}, and \eq{eq:phdynamics},
reduce to two first-order differential equations
\be
2 \L   &=& \fr{1}{2} \ls \fr{\pa W  }{\pa \ph} \rs^2 - \fr{1}{3} W^2  , \la{res:Esequaiton1}  \\
\dot{\ph} &=& - \fr{\pa W  }{\pa \ph}   . \la{res:Esequaiton2} 
\ee
These three first-order differential equations are consistent with the canonical momenta in \eq{def:conjugatemomenta} and Hamiltonian constraint in \eq{Result:HamCon}. Using these relations, the radial coordinate dependence of $\ph$ is related to the $\b$-function of the 't Hooft coupling constant
\be			\la{Result:betabulk}
\b_\ph   \equiv \fr{\pa \ph}{\pa \log \m} 
= -  \fr{N_c}{4  } \fr{\b_\l}{\l^2}   .
\ee
Solving the Hamiltonian constraint in \eq{res:Esequaiton1}, the superpotential is given by \cite{Papadimitriou:2010as,Papadimitriou:2004ap,Csaki:2006ji}
\be		\la{res:1counterterm}
W =   \fr{6 }{R}  \cosh \ls \sqrt{\fr{2}{3}} \ls \ph - \ph_0 \rs \rs .
\ee
The remaining equations determine the scalar field and metric as the following forms
\be		 
\ph - \ph_0 &=&  \et {\sqrt\fr{3}{2}} \log \ls \fr{4 \sqrt{6} - \ph_1 u^4/R^4 }{4 \sqrt{6} 
+  \ph_1 u^4/R^4 } \rs ,  \la{sol:ESeq} \\
e^{2 A(y)} &\equiv& \ls \fr{\m}{R}\rs^2=\fr{R^2}{u^2} \sqrt{1-  \fr {  \ph_1^2}{96} \fr{u^8}{R^8}} \la{Solution:gmetric} ,
\ee
with
\be			\la{met:bounarymetric}
u = R e^{- y/R} ,
\ee
where $y$ indicates the radial position of the boundary and $\ph_0$ and $\ph_1$ are two arbitrary integral constants. In this case, $\ph_0$ correspond to the coupling constant at a UV fixed point ($u \to 0$). Here we choose $\ph_1$ as a non-negative value, $\ph_1 \ge 0$, so that $\et=\pm 1$ means the sign of $\ph_1$. Relying on the value of $\et$, a marginall operator at the tree level changes into a marginally relevant or marginally irrelevant one at the quantum level.

Since the boundary action suffers from a UV divergence, as mentioned before, we need to renormalize it. Adding the following counterterm
\be             \la{Result:ctformarginal}
{\cal L}_{ct} = \fr{6 }{ R}  .
\ee
we easily check that the renormalized boundary action becomes finite even at the UV fixed point ($y \to \infty$)
\be
\G [\g_{\m\n},\ph; \bar{y}]= S_B - S_{ct} ,
\ee
which on the dual field theory side is reinterpreted as the renormalized generating functional. As a result, the RG equation of the dual field theory is governed by
\be		
0 = \fr{\m}{\sqrt{\g}} \fr{\pa \G}{\pa \m}  + \g^{ij} \bra T_{ij} \ket + \b_\ph  \bra G \ket  ,
\ee
with 
\be
\b_\ph  &=& - \fr{6 }{W} \fr{\pa W  }{\pa \ph} ,  \la{Result:Qbetafun} \\
\bra T_{ij} \ket &=&   \fr{1}{\k^2}  \ls K_{ij}  - \g_{ij} K \rs-  \fr{3}{\k^2 R}  \g_{ij}  , \\
\bra G \ket &=& \fr{1}{2  \k^2}  \fr{\pa W  }{\pa \ph} .   
\ee

Near the asymptotic region ($u \to 0$), the bulk scalar field has the following expansion
\be
\ph = \ph_0 - \fr{\et \ph_1}{4 R^4} u^4 + \cdots .
\ee 
This is consistent with \eq{Result:pscalar} for a marginal operator with $\D=4$. At $u = 0$, $\ph_0$ is identified with the coupling constant at the UV fixed point. At the tree level, a marginal operator gives rise to a vanishing $\b$-function, $\b_{tree} = 0$ in \eq{Result:treebeta}. However, the $\b$-function in \eq{Result:Qbetafun} can have a nonvanishing value. This implies that $\b_\ph$ in \eq{Result:Qbetafun} is a non-perturbative result involving all quantum corrections. To see more details, we first take $\et=1$ which leads to a marginally relevant deformation satisfying $\b_\l <0$ in \eq{Result:betabulk}. Using the superpotential in \eq{res:1counterterm}, the $\b$-function and the vev of operators in the UV region allows the following perturbative expansions
\be
\b_{\ph} &=&  \fr{\ph_1}{R^4} \fr{1}{\m^4} - \fr{\ph_1^3}{48 R^{12}} \fr{1}{\m^{12}}  + {\cal O} \ls \m^{-20} \rs         ,   \la{Result:betafuncph}  \\
\bra G \ket  &=& - \fr{\ph_1}{2 \k^2 R^5} \fr{1}{\m^4}  + {\cal O} \ls \m^{-28} \rs    , \la{Result:gluoncondensate} \\
\bra {T^{\m}}_{\m}\ket &=& - \fr{\ph_1^2}{4 \k^2 R^9} \fr{1}{\m^8} + \frac{\phi _1^4}{384 \kappa ^2   R^{17}}  \fr{1}{\m^{16}}+ {\cal O} \ls \m^{-24} \rs     .
\ee
There are several remarks we should note. First, since we consider a marginal operator, its $\b$-function must vanish at the tree level. However, $\b_\ph$ in \eq{Result:betafuncph} does not vanish. This indicates that the nonvanishing $\b_\ph$ comes from quantum (loop) effects. Second, \eq{Result:gluoncondensate} shows that the vev of the gluon condensate is proportional to $\ph_1$. Therefore, identifying $\ph_1$ with the gluon condensate in the UV region is consistent with identifying the boundary value of $\ph$ with the coupling constant relying on the RG scale. Lastly, the obtained quantities satisfy the following relation
\be			\la{Result:GRGfloweq}
\bra {T^{\m}}_{\m} \ket   =  - \fr{N_c }{8} \fr{\b_\l}{\l^2} \bra G \ket + {\cal O} \ls \l^{-4} \rs.
\ee
This up to a numerical factor is consistent with the one-loop trace anomaly \eq{Result:GRGfloweqQCD} derived in the lattice QCD \cite{DiGiacomo:1981lcx,Trinchero:2013joa}. The holographic result further shows that the trace anomaly \eq{Result:GRGfloweqQCD} is valid only at the one-loop level. Higher order quantum corrections provide additional contributions to the trace anomaly and modify the one-loop level trace anomaly, relation between the trace anomaly and gluon condensate.


\section{Conclusion}

In the present work, we have studied the holographic RG flow triggered by the marginal gluon condensate. In the lattice QCD, it was known that the gluon condensate generates a trace anomaly \cite{DiGiacomo:1981lcx,Trinchero:2013joa}. In the UV region, the trace anomaly at the one-loop level is proportional to the vev of the gluon condensate. In the traditional QFT, it is very difficult to calculate all higher order quantum effects because of infinitely many higher loop corrections. In this situation, the holographic method may be useful in understanding non-perturbative feature of QCD. According to the AdS/CFT correspondence, the non-perturbative quantum nature of a QFT can be realized by a classical gravity theory. 

To study the gluon condensate holographically, we introduced a massless scalar field in the bulk, which maps to a marginal operator. In the holographic RG flow procedure, the boundary position of a bulk geometry can be regarded as the energy scale observing the dual QFT. In this case, the boundary value of a bulk field is reinterpreted as a coupling constant coupled to the dual operator. In general, there are many marginal operators in QCD. Here, we identified a bulk field with the coupling constant of a gauge field's kinetic term to consider the gluon condensate. Then, we investigated the coupling constant depending on the energy scale by varying the boundary position of the bulk geometry. Using this holographic RG flow procedure, we reproduced the one-loop trace anomaly of the lattice QCD. In addition, we calculated higher order quantum corrections perturbatively in the UV region. We showed that higher loop corrections provide additional trace anomaly which modifies the one-loop trace anomaly known in the lattice QCD.   

In this work, we have focused on the holographic RG flow in the UV region in order to compare the holographic result with the QCD result known in the UV region. It would be also interesting and important to investigate IR behavior. However, the gravity theory considered here is IR incomplete. To investigate IR physics, we have to take into account a more general gravity theory having an IR fixed point. We hope to report some interesting results related to the holographic description of an IR physics in future works.

\vspace{1cm}

{\bf Acknowledgement}

This work was supported by the National Research Foundation of Korea(NRF) grant funded by the Korea government(MSIT) (No. NRF-2019R1A2C1006639).




%

\end{document}